\documentclass[preprint,floatfix,aps,prd,amsmath,amssymb,nofootinbib,preprintnumbers,superscriptaddress]{revtex4}

\usepackage{graphicx}
\usepackage{array}

\begin{document}

\begin{flushright}
ICRR-Report-597-2011-14\\
UT-11-35\\
\end{flushright}

\title{
 Prospects for determination of thermal history after inflation
 with future gravitational wave detectors
}

\author{Sachiko Kuroyanagi} \email[]{skuro@icrr.u-tokyo.ac.jp}
\affiliation{Institute for Cosmic Ray Research, The University of Tokyo,
  Chiba 277-8582, Japan}

\author{Kazunori Nakayama} 
\affiliation{Department of Physics, The University of Tokyo, Tokyo 113-0033, Japan}

\author{Shun Saito}
\affiliation{Department of Astronomy, University of California at Berkeley, California 94720, USA}

\date{\today}

\begin{abstract}
  Thermal history of the Universe between inflation and big-bang
  nucleosynthesis has not yet been revealed observationally.  It will
  be probed by the detection of primordial gravitational waves
  generated during inflation, which contain information on the
  reheating temperature as well as the equation of state of the
  Universe after inflation.  Based on Fisher information formalism, we
  examine how accurately the tensor-to-scalar ratio and reheating
  temperature after inflation can be simultaneously determined with
  space-based gravitational wave detectors such as the DECI-hertz
  Interferometer Gravitational-wave Observatory (DECIGO) and the
  Big-Bang Observer (BBO).  We show that the reheating temperature is
  best determined if it is around $10^7$GeV for tensor-to-scalar
  ratio of around $0.1$, and explore the detectable parameter space.
  We also find that equation of state of the early Universe can be
  also determined accurately enough to distinguish different
  equation-of-state parameters if the inflationary gravitational waves
  are successfully detected.  Thus future gravitational wave detectors
  provide a unique and promising opportunity to reveal the thermal
  history of the Universe around $10^7$GeV.
\end{abstract}


\maketitle

\section{Introduction}
The standard paradigm in cosmology is well established and explains
cosmological evolution after the big-bang nucleosynthesis (BBN).  An
epoch of primordial inflation at the very early Universe now also
becomes an essential part of the standard paradigm, and is supported
by accurate measurements of the cosmic microwave background (CMB)
anisotropies.  The nearly scale-invariant power spectrum of the
curvature perturbation, as well as its Gaussian nature, strongly
indicates the inflation as a source of the density fluctuation of the
Universe \cite{Komatsu:2010fb}.

However, thermal history {\it between} inflation and BBN has not yet
revealed observationally.  There must be an epoch of reheating after
inflation, where the inflaton decays and the radiation-dominated
universe starts.  The process of reheating crucially depends on the
properties of the inflaton: its mass, potential and couplings to
standard model particles.  Before the inflaton decays, inflaton
oscillates around its potential minimum.  In the simplest class of
inflation models, the inflaton oscillation behaves as matter after
inflation ends.  But the detail of the inflaton oscillation, or the
equation of state of the Universe during the inflaton oscillation
depends on the nature of the inflaton.  Therefore, it is extremely
important to probe the thermal history after inflation.  It would
determine the nature of the inflaton and might be a direct hint for
the underlying high-energy theory such as supergravity or string
theory.

Then, how can we probe the very early Universe observationally?
Perhaps gravitational waves may be a unique signal that directly
carries information on the thermal history of these early epochs,
since the Universe is transparent to gravitational waves up to the
Planck epoch in principle.  Fortunately, there is a promising source
of the gravitational wave: primordial gravitational waves generated by
inflation \cite{Starobinsky:1979ty}.  Quantum fluctuations during
inflation are expanded and frozen outside the Hubble horizon.  Tensor
fluctuations generated in this way can be observed as gravitational
waves.  These gravitational waves form a stochastic background having
a very wide range of wavelength from the present horizon scale to the
terrestrial one.  If the inflationary energy scale is sufficiently
large, inflationary gravitational waves will be detected through
on-going or future CMB B-mode polarization measurements
\cite{Kamionkowski:1996ks,Saito:2007kt} and/or space laser
interferometers \cite{Turner:1993vb,Smith:2005mm}.  Proposed future
gravitational wave measurements such as the DECI-hertz Interferometer
Gravitational-wave Observatory (DECIGO)
\cite{Seto:2001qf,Kawamura:2006up} and the Big-Bang Observer (BBO)
\cite{bbo,Cutler:2009qv} aim at the detection of the inflationary
gravitational wave background around $0.1$-$1\,$Hz as well as the
determination of the nature of the present accelerated expansion of
the Universe with binary sources as standard sirens
\cite{Cutler:2009qv,Nishizawa:2010xx,Nishizawa:2011eq}.

Interestingly, the spectrum of the inflationary gravitational wave
background directly reflects the expansion history of the Universe
\cite{Seto:2003kc,Tashiro:2003qp,Boyle:2005se,Boyle:2007zx}.  This is
because the gravitational wave amplitude frozen at super-horizon scale
begins to damp inversely proportional to the scale factor of the
Universe when the corresponding mode enters the horizon.  As we have
already discussed, the matter-dominated reheating period is followed
by the radiation-dominated era.  Therefore, the spectrum shows a
knee-like feature at the frequency corresponding to the comoving
horizon scale at the end of reheating, which is given by $f_{\rm R}
\simeq 0.26\,{\rm Hz} (T_{\rm R}/10^7\,{\rm GeV})$ (see
Eq.~(\ref{f_R}) below) where $T_{\rm R}$ is the temperature of the
Universe at the end of reheating defined by
\begin{equation}
	T_{\rm R} = \left( \frac{10}{\pi^2 g_{*}(T_{\rm R})} \right)^{1/4}\sqrt{ \Gamma_\phi M_P },
\end{equation}
with $\Gamma_\phi$ being the inflaton decay rate and $M_P$ being the
reduced Planck scale.  If future gravitational wave detectors see the
knee shape at this frequency, it will be a direct measurement of the
reheating temperature.

The prospects for direct detection of the inflationary gravitational
wave background with the effect of reheating have been investigated in
Refs. \cite{Nakayama:2008ip,Nakayama:2008wy,Kuroyanagi:2010mm}.  The
signal-to-noise ratio is calculated taking into account the effect
that reheating induces a change of the frequency dependence in the
spectrum, and the detectability of the gravitational wave background
is discussed in the previous works.  However, direct detection of the
gravitational wave amplitude does not necessarily mean accurate
determination of the reheating temperature.  In order to estimate the
detectability of the reheating signature more realistically, we need
to examine if the experiment can distinguish the change of the
characteristic frequency dependence induced by reheating.  For that
purpose, we employ the Fisher matrix formalism, which enables us to
evaluate the sensitivity to the shape of the spectrum with properly
taking into account parameter degeneracies.

As we have mentioned, the change of the frequency dependence is
attributed to the fact that reheating causes the change of the Hubble
expansion rate.  Thus, the inflationary gravitational wave background
is useful not only determining the reheating temperature but also
constraining the expansion history of the very early Universe.  We
evaluate the ability of future direct detection experiments to
determine the equation of state of the early Universe, keeping in mind
that many applications on cosmological models are expected with direct
detection of the inflationary gravitational wave background
\cite{Seto:2003kc,Tashiro:2003qp,Boyle:2005se,Boyle:2007zx,Nakayama:2008ip,
  Nakayama:2008wy,Kuroyanagi:2008ye,Nakayama:2009ce,Nakayama:2010kt,
  Mukohyama:2009zs,Kuroyanagi:2010mm}.

In this paper, we study prospects for determination of the reheating
temperature, as well as the equation of state of the early Universe,
at future gravitational wave detectors based on the Fisher matrix
method.  In Sec.~\ref{sec:Fisher}, basic formulations are provided.
In Sec.~\ref{sec:TR}, we derive the accuracy of the reheating
temperature measurement with future gravitational wave detectors.  In
Sec.~\ref{sec:w}, we study how accurately the equation of state of the
early Universe will be determined in the same setup.  The final
section is devoted to conclusions.  The fiducial cosmological
parameters are taken to be the maximum likelihood values from the
combined analysis of the WMAP 7-year, baryon acoustic oscillation, and
supernova data with a flat $\Lambda$CDM universe
\cite{Komatsu:2010fb}: matter density $\Omega_m h^2=0.1344$, amplitude
of curvature perturbation $\Delta_{{\cal R},{\rm prim}}^2=2.45\times
10^{-9}$ and the Hubble parameter $h=0.702$.  Throughout the paper we
choose the natural units, $c=\hbar=1$.

\section{Fisher matrix method} 
\label{sec:Fisher}
\subsection{Definitions of gravitational wave observables}

Before discussing the Fisher matrix approach, let us first summarize
the observable quantities of the gravitational wave detectors.
Gravitational waves in the expanding Universe are described as tensor
perturbations in the Friedmann-Robertson-Walker metric,
$ds^2=a^2(t)[-d\tau^2+(\delta_{ij}+h_{ij})dx^idx^j]$, with $a(t)$
being the scale factor of the Universe.  The conformal time $\tau$ is
defined as $d\tau\equiv dt/a(t)$, and the subscript $i$ and $j$ denote
spatial indices which run over $1,2,3$ or $x,y,z$.  The tensor
perturbation $h_{ij}$ satisfies the transverse-traceless conditions,
$\partial^ih_{ij}=h^i_i=0$.  It is expanded into its Fourier series as
\begin{equation}
    h_{ij}(t,\textbf{x})=\sum_{\lambda=+,\times}^{}\int\frac{d^3k}{(2\pi)^{3/2}}\epsilon_{ij}^{\lambda}
    (\textbf{k})h_\textbf{k}^{\lambda}(t)e^{i\textbf{k}\cdot\textbf{x}},
\end{equation}
where the polarization tensors $\epsilon_{ij}^{+,\times}$ satisfy
symmetric and transverse-traceless conditions and are normalized as
$\sum_{i,j}^{}\epsilon_{ij}^{\lambda}(\epsilon_{ij}^{\lambda^{\prime}})^*=
2\delta^{\lambda\lambda^{\prime}}$.  Then the observable intensity of
a stochastic gravitational wave background is characterized by its
density parameter per logarithmic wavenumber (or frequency) as
\begin{equation}
     \Omega_{\rm GW}\equiv\frac{1}{\rho_c}\frac{d\rho_{\rm GW}}{d\ln k}
     =\frac{1}{12}\left(\frac{k}{aH}\right)^2\frac{k^3}{\pi^2}
     \sum_{\lambda}^{}|h_\textbf{k}^{\lambda}|^2,
\end{equation}
where $\rho_{\rm GW}$ denotes the energy density of the gravitational
waves, $\rho_{\rm GW}=\langle (\partial_{\tau}h_{ij})^2+(\vec{\nabla}
h_{ij})^2 \rangle/(64\pi Ga^2)$, $\rho_{c}\equiv3H^2/8\pi G$ is the
critical density of the Universe, and the Hubble parameter is defined
as $H\equiv(da/dt)/a$.  One can also express such a statistical
quantity in terms the power spectrum,
\begin{equation}
     \Delta_h^2(k)\equiv\frac{d\langle h_{ij}h^{ij}\rangle}{d\ln k}
     =\frac{k^3}{\pi^2}\sum_{\lambda}^{}|h_\textbf{k}^{\lambda}|^2.
\end{equation}

\subsection{Fisher matrix for gravitational wave measurement}

Here, we briefly review and discuss the Fisher information matrix
approach, focusing on the future gravitational wave detectors.  The
Fisher matrix is a powerful method to theoretically forecast the
constraining power on parameters of interest for a given survey, and
is commonly used in observational cosmology (see e.g.,
\cite{Tegmark:1996bz,Eisenstein:1998hr,Saito:2009ah} ).  An essential
but unique assumption in the formalism is a Gaussian likelihood, and
the Fisher matrix is defined by the second derivative of the
likelihood around its maximum (or fiducial parameters) with respect to
the parameters of interest. Then the inverse of the Fisher matrix
provides a lower bound on the covariance matrix via the Cramer-Rao
bound, and is regarded as the best achievable accuracy for the
parameters.

The Fisher matrix generally depends on the covariance matrix of
signals, i.e., the noise properties of the survey configurations.  In
the case of the stochastic gravitational wave background, direct
detection can be attempted by cross-correlating the output signals
between two detectors \cite{Allen:1997ad}.  For the cross-correlation
analysis, the Fisher information matrix is given by \cite{Seto:2005qy}
\begin{equation}
     {\cal F}_{ij}=\left(\frac{3H_0^2}{10\pi^2}\right)^2 2\,T_{\rm obs}\nonumber\\
     \sum_{(I,J)}\int^{f_{\rm max}}_{f_{\rm cut}}df\frac{|\gamma_{IJ}(f)|^2\partial_{p_i}
     \Omega_{\rm GW}(f)\partial_{p_j}\Omega_{\rm GW}(f)}{f^6S_I(f)S_J(f)},
     \label{Fisher}
\end{equation}
with $f=k/(2\pi)$ and $H_{0}$ being the frequency of the gravitational
wave and the present Hubble parameter.  We choose a lower cutoff of
$f_{\rm cut}=0.1$Hz, below which the signal may be contaminated by
noise from cosmological white dwarf binaries \cite{Farmer:2003pa}.  As
for $f_{\rm max}$, we set $f_{\rm max}=\infty$ unless otherwise
stated, though the high-frequency range is limited by the noise
spectrum which we will see in detail later.  For a given survey, we
need to assign three observational ingredients; a total observational
$T_{\rm obs}$, overlap reduction functions $\gamma_{IJ}(f)$, and noise
spectra $S_{I}(f)$.  These functional forms rely on the type of
interferometry that we choose (for a review, see
\cite{Maggiore:1900zz}).

In the case of the Time-Delay Interferometry (TDI) that is expected to
be adopted in BBO, the subscript $I$ or $J$ denotes the TDI channel
output index ($I,J=A,E,T$).  We compute the overlap reduction function
$\gamma_{IJ}(f)$ for the TDI data combinations with the method in Ref.
\cite{Corbin:2005ny}.  The noise transfer functions for the TDI
variables are assumed as \cite{Prince:2002hp}
\begin{equation}
     S_A(f)=S_E(f)=8\sin^2(\hat{f}/2)[(2+\cos\hat{f})S_{\rm shot}
     +2(3+2\cos\hat{f}+\cos(2\hat{f}))S_{\rm accel}],
     \label{eq: S_AE}
\end{equation}
\begin{equation}
     S_T(f)=2[1+2\cos\hat{f}]^2[S_{\rm shot}+4\sin^2(\hat{f}/2)S_{\rm accel}],
     \label{eq: S_T}
\end{equation}
where $S_{\rm shot}$ and $S_{\rm accel}$ are the photon shot-noise and
the proof-mass acceleration noise for the laser interferometers.  The
noise spectrum is designed so as to be rescaled with respect to the
pivot frequency $\hat{f}=2\pi Lf$ with $L$ being the arm length.
Unlike BBO, DECIGO would install a Fabry-Perot type interferometer,
and hence its noise functions can be different from those of BBO
discussed above.  However, both DECIGO and BBO are basically designed
to aim at the detection of the inflationary gravitational waves with
similar frequency ranges around $0.1-1$Hz.  In addition, the purpose
of this paper is to quantify the potential of future gravitational
wave experiments for determination of model parameters.  Based on
these facts, we simply assume the TDI-type noise functions,
Eqs.~(\ref{eq: S_AE}) and (\ref{eq: S_T}), throughout the paper.  Even
with such a simple assumption, we do not believe that detailed
configurations can significantly affect our conclusion.

In table~\ref{table: survey params}, we summarize the survey
parameters adopted in the Fisher matrix calculation.  We consider two
types of detectors.  The first one is corresponding to the proposed
DECIGO or BBO, quoted by "BBO/FP-DECIGO", whose parameter values are
taken from Ref. \cite{Crowder:2005nr}.  We also investigate the case
of an ideal experiment whose sensitivity is limited only by quantum
noises.  We quote it as "Ultimate-DECIGO" and its parameters are taken
from Ref.  \cite{Kudoh:2005as}.
\begin{table}[t]
     \begin{tabular}{c|ccc}
     ~Detectors~ 
     & \shortstack{$S_{\rm shot}$ \\ ~$[(L/{\rm km})^{-2}{\rm Hz}^{-1}]$~ }
     & \shortstack{$S_{\rm accel}$ \\ ~$[(2\pi f/{\rm Hz})^{-4}(
2 
L/{\rm km})^{-2}{\rm Hz}^{-1}]$~}
     & \shortstack{~$L$~ \\ ~[km]~ }\\
     \hline
     \hline
     ~BBO/FP-DECIGO~
     & ~$2\times 10^{-40}$~
     & ~$9\times 10^{-40}$~
     & ~$5\times 10^{4}$~\\
     \hline 
     ~Ultimate DECIGO~
     & ~$9\times 10^{-44}$~
     & ~$9\times 10^{-44}$~
     & ~$5\times 10^{4}$~\\
     \end{tabular}
     \caption{detector parameters}
      \label{table: survey params}
\end{table}

In summary, the Fisher matrix can be calculated with
Eq.~(\ref{Fisher}), once the detector parameters are assigned and
theoretical predictions for $\Omega_{\rm GW}$ is provided.  Then the
marginalized $1\sigma$ error is easily computed with the inverse of
the Fisher matrix,
\begin{equation}
     \sigma(p_{i}) =\sqrt{({\cal F}^{-1})_{ii}}. 
\end{equation}
The Fisher matrix is a product of the signal-to-noise-ratios,
$\Omega_{\rm GW}/S$, and the derivative, $\partial_{p_{i}}\ln
\Omega_{\rm GW}$, and hence depends on the parameter response, namely
the parameter degeneracy as well as the signal detectability.  This
makes a difference from the previous work \cite{Nakayama:2008wy}.

\section{Determination of the reheating temperature}  
\label{sec:TR}
As discussed in
\cite{Nakayama:2008ip,Nakayama:2008wy,Kuroyanagi:2010mm}, direct
detection of the inflationary gravitational wave background has a
potential to constrain or even to determine the reheating temperature
via the characteristic frequency dependence. Inflationary
gravitational wave spectrum has a frequency dependence of $f^{-2}$ for
modes which enter the horizon during a matter-dominated universe, and
$f^0$ for modes which enter the horizon during a radiation-dominated
universe.  If the Universe behaves like a matter-dominated universe
during reheating, the transition from reheating to the radiation
domination is seen as a change of the frequency dependence of the
spectrum.  If this signature exists in the frequency band of direct
detection sensitivity, $0.1-1$Hz, we may be able to determine the
reheating temperature by measuring the knee-like feature where the
frequency dependence changes from $f^{-2}$ to $f^0$.  In this section,
we apply the Fisher matrix analysis to investigate to what extent the
future gravitational wave experiments can determine the reheating
temperature.

\subsection{The gravitational wave spectrum in the presence of reheating}
\begin{figure}
 \begin{center}
  \includegraphics[width=0.6\textwidth]{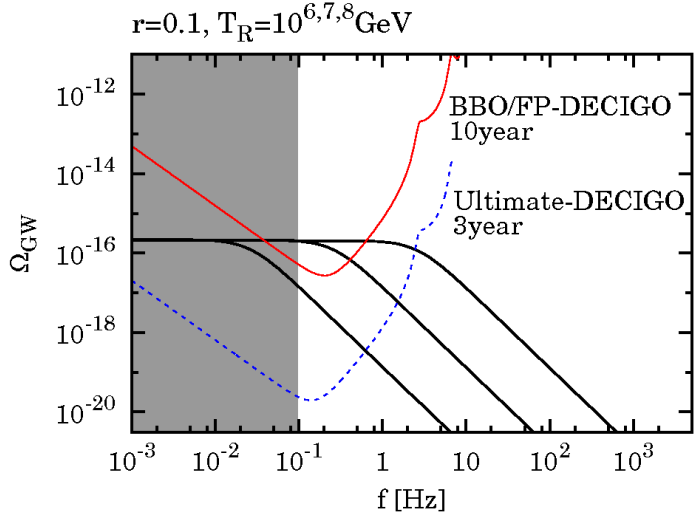}
  \caption{
The spectra of the inflationary gravitational wave background for
different values of the reheating temperature (thick black solid
curves with $T_{\rm R}=10^{6,7,8}\,$GeV from left to right).  The
tensor-to-scalar ratio is taken to be $r=0.1$.  For reference, the
noise spectra for BBO/FP-DECIGO with 10-year observation (red solid)
and for Ultimate-DECIGO with 3-year observation (blue dotted) are
shown.  The gray shaded region is not used in the Fisher analysis,
since noises from white dwarf binaries may significantly contribute as
systematic errors.
    \label{fig: spectrum}
  }
 \end{center}
\end{figure}

The spectrum of the inflationary gravitational wave background is
often expressed in terms of the initial tensor power spectrum
$\Delta_{h,{\rm prim}}^2(k)$ and the transfer function $T_h(k)$,
\begin{equation}
     \Omega_{\rm GW}=\frac{1}{12}\left(\frac{k}{aH}\right)^2 \Delta_{h,{\rm prim}}^2(k)T_h^2(k).
     \label{OmegaGW1}
\end{equation}
In a single field slow-roll inflation, the tensor-to-scalar ratio
$r\equiv \Delta_{h,{\rm prim}}^2(k_0)/\Delta_{{\cal R},{\rm
    prim}}^2(k_0)$ can be related to the tilt of the tensor mode
spectrum $n_T\equiv d\ln \Delta_{h,{\rm prim}}^2(k_0)/d\ln k$ as
$r=-8n_T$.  From this relation, the initial power spectrum can be
written as
\begin{equation}
     \Delta_{h,{\rm prim}}^2(k)\simeq r\Delta_{{\cal R},{\rm prim}}^2(k_0)
     \exp{\left[-\frac{r}{8}\ln\frac{k}{k_0}+\cdots\right]},
     \label{primordial}
\end{equation}
where the pivot scale is taken as $k_0=0.002\,{\rm Mpc}^{-1}$.
\footnote{ The contribution from the higher order of $\ln(k/k_0)$ is
  sometimes non-negligible, depending on the inflation model, and may
  cause a wrong estimate of the amplitude of the spectrum
  \cite{Kuroyanagi:2011iw}.  It can affect determination of the value
  of $r$, which is an important parameter to determine the
  gravitational wave amplitude, but is not crucial for determination
  of the reheating temperature, because the reheating temperature is
  determined basically by the characteristic frequency dependence of
  the spectrum.  } The effects of the cosmological evolution after
inflation are all included in the transfer function, which is given as
\cite{Nakayama:2008wy,Nakayama:2009ce}
\begin{equation}
     T_h^2(k)=\Omega_m^2\left(\frac{g_*(T_{\rm in})}{g_{*0}}\right)
     \left(\frac{g_{*s0}}{g_{*s}(T_{\rm in})}\right)^{4/3}
     \left(\frac{\overline{3j_1(k\tau_0)}}{k\tau_0}\right)^2
     T_1^2(x_{\rm eq})T_2^2(x_{\rm R}).
     \label{transfer}
\end{equation}
The subscript ``0'' denotes the present time and ``in'' denotes the
time when the mode $k$ crosses the horizon.  The effective number of
degrees of freedom at the end of reheating is taken to be the sum of
the standard model particles, $g_*(T_{\rm R})=g_{*s}(T_{\rm
  R})=106.75$.  The values at present are $g_{*0}=3.36$ and
$g_{*s0}=3.90$.  Here, $\tau_0$ is the present conformal time
calculated assuming the Universe is matter dominated:
$\tau_0=2H_0^{-1}$.  The effect of the cosmological constant is
accounted for by the factor of $\Omega_m=1-\Omega_{\Lambda}.$ In the
limit of $k\tau_0\ll 1$, the spherical Bessel function $j_1(x)=(\sin
x-x\cos x)/x^2$ is replaced as $\overline{j_1(k\tau_0)}\rightarrow
\overline{\cos(k\tau_0)}/(k\tau_0)=1/(\sqrt{2}k\tau_0)$.  The first
transfer function $T_1(x_{\rm eq})$ describes the change of the
frequency dependence of the spectrum which arises from the change of
the expansion rate of the Universe at the matter-radiation equality
$t=t_{\rm eq}$~\cite{Turner:1993vb},
\begin{equation}
     T_1^2(x_{\rm eq})=(1+1.57x_{\rm eq}+3.42x_{\rm eq}^2),
     \label{trans1}
\end{equation}
where $x_{\rm eq}=k/k_{\rm eq}$ and $k_{\rm eq}\equiv\tau_{\rm
  eq}^{-1}=7.1\times 10^{-2}\Omega_m h^2\,{\rm Mpc}^{-1}$.  The second
transfer function $T_2(x_{\rm R})$ corresponds to the change of the
expansion rate at the end of reheating $t=t_{\rm
  R}$~\cite{Nakayama:2008wy},
\begin{equation}
     T_2^2(x_{\rm R})=(1-0.32x_{\rm R}+0.99x_{\rm R}^2)^{-1},
     \label{trans2}
\end{equation}
where $x_{\rm R}=k/k_{\rm R}$ and $k_{\rm R}\simeq 1.7\times
10^{14}{\rm Mpc}^{-1}(g_{*s}(T_{\rm R})/106.75)^{1/6}(T_{\rm
  R}/10^7{\rm GeV})$.  This can be rewritten in terms of frequency as
\begin{equation}
     f_{\rm R}=\frac{k_{\rm R}}{2\pi}\simeq 0.26{\rm Hz}
     \left(\frac{g_{*s}(T_{\rm R})}{106.75}\right)^{1/6}
     \left(\frac{T_{\rm R}}{10^7{\rm GeV}}\right),   
     \label{f_R}
\end{equation}
which is the frequency where the change of the frequency dependence
due to reheating arises.  We show the spectra for different values of
the reheating temperature in Fig.~\ref{fig: spectrum}.  As clearly
seen from the figure, the knee shape around $f_{R}$ can be observed
with BBO/FP-DECIGO if the observational time is sufficiently long.
Since there are no observable to probe reheating so far, the future
gravitational wave experiments may provide us an unique opportunity to
reveal the reheating of the Universe.

\subsection{Result}
Based on the theoretical prediction presented above, let us estimate
the detectability of the reheating temperature using the Fisher
matrix, which is calculated by substituting
Eqs.~(\ref{OmegaGW1})-(\ref{trans2}) into Eq.~(\ref{Fisher}).  We take
$r$ and $T_{R}$ as free parameters, which correspond to the amplitude
of the spectrum and the frequency of the reheating signature.

\begin{figure}
 \begin{center}
  \includegraphics[width=0.7\textwidth]{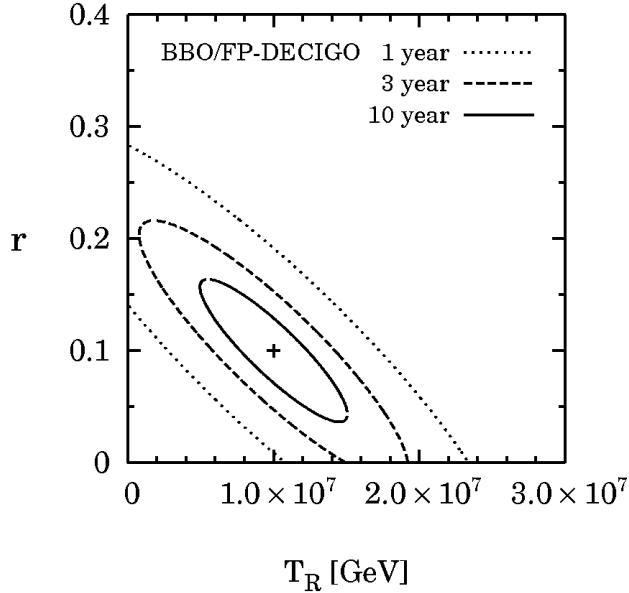}
  \caption{\label{figcontTR}The $2\sigma$ confidence level contours in
    the $T_{\rm R}-r$ plane for 1-year (dotted), 3-year
    (dashed) and 10-year (solid) observation by BBO/FP-DECIGO. The
    fiducial parameters are set as $r=0.1$ and $T_{\rm R}=10^7$GeV,
    which is shown by a cross mark.}
 \end{center}
\end{figure}

In Fig. \ref{figcontTR}, we present an example of the expected future
constraints, in which the fiducial parameters are chosen as $r=0.1$
and $T_{\rm R}=10^7\,$GeV.  Each ellipse represents the 2$\sigma$
error contours expected from 1, 3 and 10 years of observation with
BBO/DECIGO.  The error ellipse shrinks more for longer observations
due to the fact that the signal-to-noise ratio scales as $\sqrt{T_{\rm
    obs}}$.  As naturally expected, there is a degeneracy between
$T_{\rm R}$ and $r$. This is simply because, as long as the frequency
dependence of the spectrum is measured with a good accuracy we cannot
distinguish the spectrum with larger $T_{R}$ from that with smaller
amplitude, i.e., with smaller $r$.

\begin{figure}
 \begin{center}
  \includegraphics[width=0.5\textwidth]{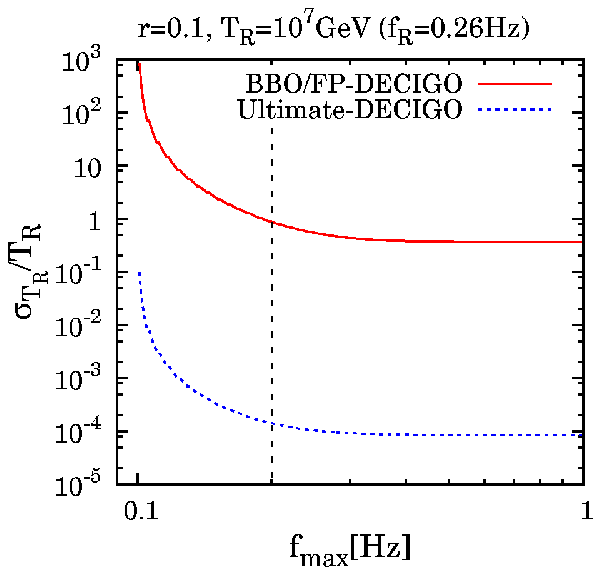}
  \caption{
The $1\sigma$ marginalized errors are shown as a function of $f_{\rm max}$, calculated assuming $r=0.1$ and $T_{\rm R}=10^{7}$GeV (corresponding to $f_{R}=0.26\,$Hz) with 3-year observations. For clarity, the best-sensitivity frequency is plotted as a vertical dotted line. 
    \label{fig: fmax}
  }
 \end{center}
\end{figure}

An interesting and nontrivial question is what frequency range
actually carries information on the reheating temperature. In other
words, how wide of a band width is necessary to detect the knee shape
with a good accuracy.  We study this issue in Fig.~\ref{fig: fmax}, in
which errors in $T_{\rm R}$ are plotted as a function of the upper
frequency limit in the calculation of the Fisher matrix ($f_{\rm max}$
in Eq. (\ref{Fisher})).  Apparently frequencies above $f\simeq
0.3\,$Hz do not contribute to detection of the reheating temperature.
This is because both the suppression of the signal amplitude due to
reheating and the increase of the noise spectrum intensity prevents us
from reaching the spectrum information.  Thus, a moderate kink around
$f_{R}$ is enough to reveal the reheating signature.

Note that our analysis is performed imposing the consistency relation,
$r=-8n_T$.  Since the tilt of the spectrum is defined at the CMB
scale, it can largely affect the amplitude at the direct detection
scale as seen in Eq. (\ref{primordial}).  Therefore, an additional
degeneracy between $r$ and $n_T$ arises and causes larger
uncertainties in parameters if we take $n_T$ as a free parameter.  For
instance, in the case of $r=0.1$ and $T_{\rm R}=10^7\,$GeV with
BBO/FP-DECIGO for 3-year observation, the uncertainty on the reheating
temperature is degraded to $\sigma_{T_{\rm R}}=2.9\times 10^7$ GeV by
a factor of 8 compared to $\sigma_{T_{\rm R}}=3.7\times 10^6$ GeV with
fixed $n_{T}$.  However, combining with CMB B-mode polarization
constraints would help to relax the degradation \cite{Powell:2011ue}.

\begin{figure}
 \begin{center}
  \includegraphics[width=0.45\textwidth]{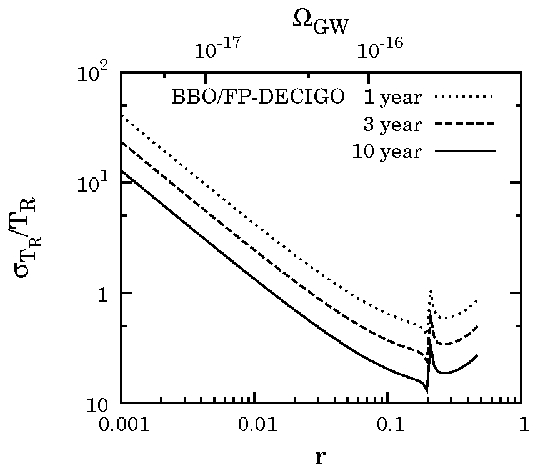}
  \includegraphics[width=0.45\textwidth]{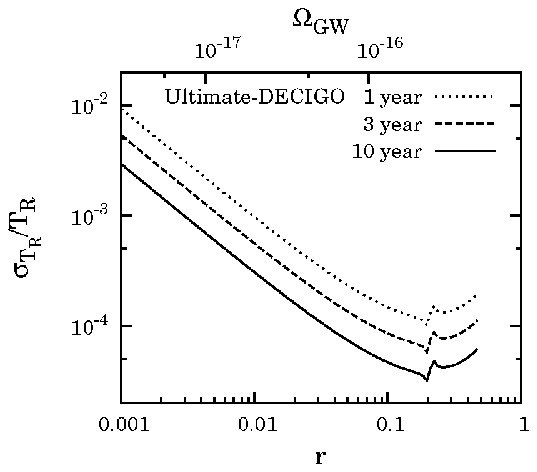}
  \caption{\label{figsigmaTR1} The marginalized $1\sigma$ uncertainty
    in $T_{\rm R}$ as a function of $r$ for BBO/FP-DECIGO (left panel)
    and Ultimate-DECIGO (right panel).  The fiducial value of the
    reheating temperature is fixed to be $T_{\rm R}=10^7$GeV.  The upper
    horizontal axis represents the values of $\Omega_{\rm GW}$
    corresponded to $r$ by Eq. (\ref{OmegaGW1}).}
 \end{center}
\end{figure}

So far, we have fixed fiducial values of $r$ and $T_{\rm R}$. In the
following, we discuss the fiducial-value dependence and predict the
parameter space where the signature of reheating can be successfully
detected.  Fig. \ref{figsigmaTR1} shows dependence on fiducial value
of $r$.  The marginalized error in $T_{\rm R}$ ($\sigma_{T_{\rm R}}$)
is calculated by changing $r$ with the fixed value of $T_{\rm
  R}=10^7$GeV.  The error becomes smaller as the gravitational wave
background is detected with larger signal-to-noise ratio.  According
to Eq.~(\ref{primordial}), the amplitude of the spectrum at the
frequency band of the experiment $f\sim 0.1$Hz takes the maximum value
when $r\simeq 0.2$ in the balance between the factors $r$ (which
increases the amplitude) and $\exp[-r\ln(k/k_0)/8]$ (which decreases
the amplitude).  This results in the smallest error on $T_{\rm R}$
around $r\simeq 0.2$ in Fig. \ref{figsigmaTR1}.  Notice that the
spikes originate from the fact that the Fisher matrix ${\cal
  F}_{rr}\propto (\partial\Omega_{\rm GW}/\partial r)^2 \propto
\{1-r\ln(k/k_0)/8\}$ goes to zero around $r\simeq 0.23$.  This is an
artificial effect due to our choice of parametrization.

\begin{figure}
 \begin{center}
  \includegraphics[width=0.45\textwidth]{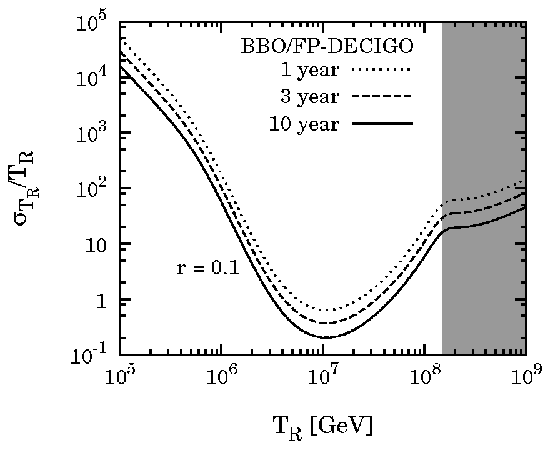}
  \includegraphics[width=0.45\textwidth]{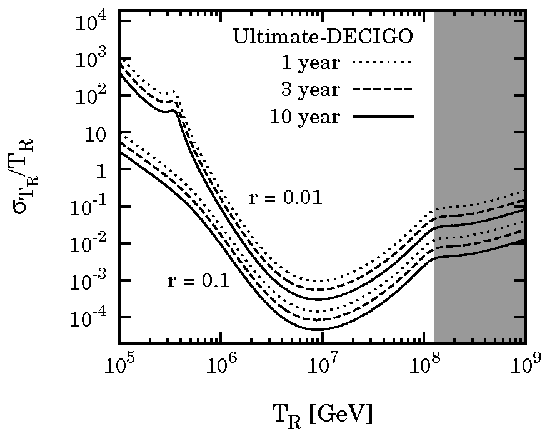}
  \caption{\label{figsigmaTR2} The marginalized $1\sigma$ uncertainty
    in $T_{\rm R}$ as a function of $T_{\rm R}$ for
    BBO/FP-DECIGO (left panel) and Ultimate-DECIGO (right panel).  The
    fiducial value of the tensor-to-scalar ratio is fixed to be $r=0.1$.}
 \end{center}
\end{figure}

Similarly in Fig.~\ref{figsigmaTR2} we show dependence on the fiducial
value of $T_{\rm R}$ with the fixed value of $r=0.1$.  The error
becomes smaller when the signature of reheating comes into the range
of the sensitivity, which corresponds to the reheating temperature of
about $10^6\,$GeV to $10^8\,$GeV.  We should comment on our results at
high $T_R$ where the resultant $\sigma_{T_{R}}$ is underestimated.
This is because the term of $-0.32x_R$ in our approximated transfer
function, Eq. (\ref{trans2}), causes a small but an artificial
``bump'' in the spectrum, which does not arise in the full numerical
calculation.  For this reason, we highlight such questionable results
at $T_R>10^{8}\,$GeV as gray-shaded regions.  The right panel of
Fig.~\ref{figsigmaTR2} shows a very promising fact in the case of
Ultimate-DECIGO that the reheating temperature could be determined
with $1\%$ accuracy if $1.2\times 10^6\,{\rm GeV}<T_{\rm R}<3.3\times
10^8\,{\rm GeV}$ \footnote{ The upper value may be smaller by a
  several factor since the value is evaluated in the shaded region of
  the figure.  }  for $r=0.1$, and if $2.1\times 10^6{\rm GeV}<T_{\rm
  R}<7.0\times 10^7{\rm GeV}$ for $r=0.01$.

\begin{figure}
 \begin{center}
  \includegraphics[width=0.45\textwidth]{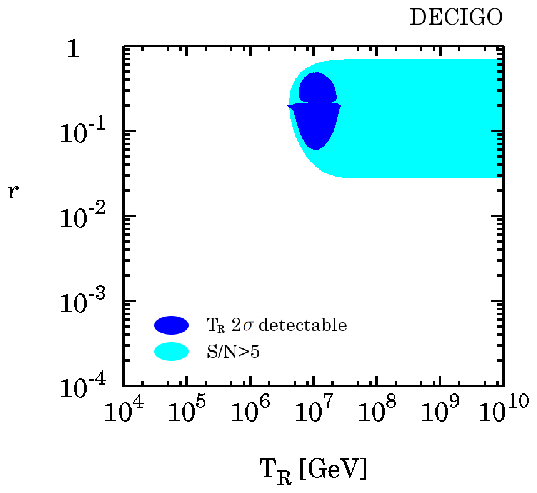}
  \includegraphics[width=0.45\textwidth]{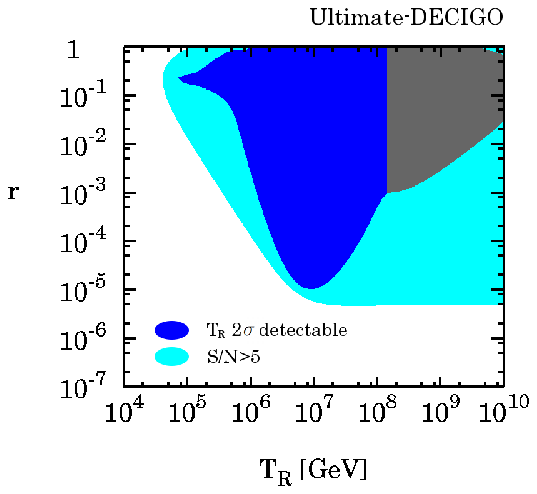}
  \caption{\label{figcontour} $2\sigma$ detection region of $T_{\rm
      R}$ is shown as a blue shaded region for 3-year observations by
    BBO/FP-DECIGO (left panel) and Ultimate-DECIGO (right panel).  The
    gray area represents the region where $\sigma_{T_{\rm R}}$ may be
    underestimated.  In the light blue shaded region, the inflationary
    gravitational wave background would be detected with a signal to
    noise ratio higher than 5. }
 \end{center}
\end{figure}

Finally in Fig. \ref{figcontour}, we present the parameter space of
$r$ and $T_{\rm R}$ where the reheating signature is detected at
greater than 2$\sigma$ level (i.e., $T_{\rm R}/\sigma_{T_{\rm R}}>2$)
with 3-year observation.  Similarly to Fig.~\ref{figsigmaTR2}, the
region affected by our incorrect transfer function are indicated by
the gray shade, and may shrink in a more realistic case.  We also show
the parameter region for the detection with a signal-to-noise ratio
higher than 5 (S/N$>5$) that is the criterion adopted in the previous
work \cite{Nakayama:2008wy}.  We cannot instantly compare two areas in
the sense that the areas largely depend on variant criteria.  However,
the shape of parameter regions are at least different, originating
from the fact parameter degeneracies are neglected in the previous
work.  We conclude that the parameter space presented here are more
realistic and worth while being pursued for detection of the reheating
temperature with future gravitational wave detectors.

\section{Probing the equation of state of the early Universe}   
\label{sec:w}
In the previous section, we have focused on the determination of the
reheating temperature by observing the change of the frequency
dependence of the spectrum.  On another front, direct detection of the
inflationary gravitational wave background can be used to determine
the equation of state of the very early Universe, at the cosmic
temperature of around $10^7\,$GeV.  So we could ask how sensitive the
detectors are to different equation-of-state parameters.  Here, we
calculate the Fisher matrix assuming that the frequency dependence of
the spectrum is uniform over the range of sensitivity, and investigate
how accurately direct detection can determine the equation of state of
the early Universe.

\subsection{The spectrum with a generic equation of state}
Let us first relate the equation of state of the Universe $w=p/\rho$
to the tilt of the gravitational wave background spectrum.  If the
initial power spectrum is assumed to have no tilt, $\Delta_{h,{\rm
    prim}}^2\propto k^0$, the frequency dependence of the spectrum is
determined only by the transfer function.  Hence, Eq. (\ref{OmegaGW1})
implies $\Omega_{\rm GW}\propto k^2 T_T^2(k)$.  Since a gravitational
wave, which has an initial amplitude of $h_{\textbf{k},{\rm
    prim}}^{\lambda}$, maintains constant amplitude outside the
horizon and it starts to decrease inversely proportional to the scale
factor when the mode enters the horizon, the transfer function 
can be written as $T_T(k)=|h_{\textbf{k},0}|/|h_{\textbf{k},\rm
  prim}|=(a_0/a_{\rm in})^{-1}$, which means $\Omega_{\rm GW}\propto
k^2 a_{\rm in}^2$.  Combining the facts that the mode enter the
horizon when $k=aH$ and the Hubble expansion rate is given in terms of
the equation of state as $H^2\propto a^{-3(1+w)}$, we obtain $a_{\rm
  in}\propto k^{-2/(1+3w)}$.  Thus, for modes which enter the horizon
when the Universe has the equation of state $w$, the spectrum has the
frequency dependence of
\begin{equation}
\Omega_{\rm GW}\propto k^{\frac{2(3w-1)}{3w+1}}.
\end{equation}
We parametrize the amplitude of the gravitational wave
background spectrum, normalizing at $f=F$, as
\begin{equation}
\Omega_{\rm GW}(f)=\Omega_{\rm GW,F}(f/F)^{\frac{2(3w-1)}{3w+1}}.
\label{OmegaGW2}
\end{equation}

\subsection{Result}
\begin{figure}
 \begin{center}
  \includegraphics[width=1.0\textwidth]{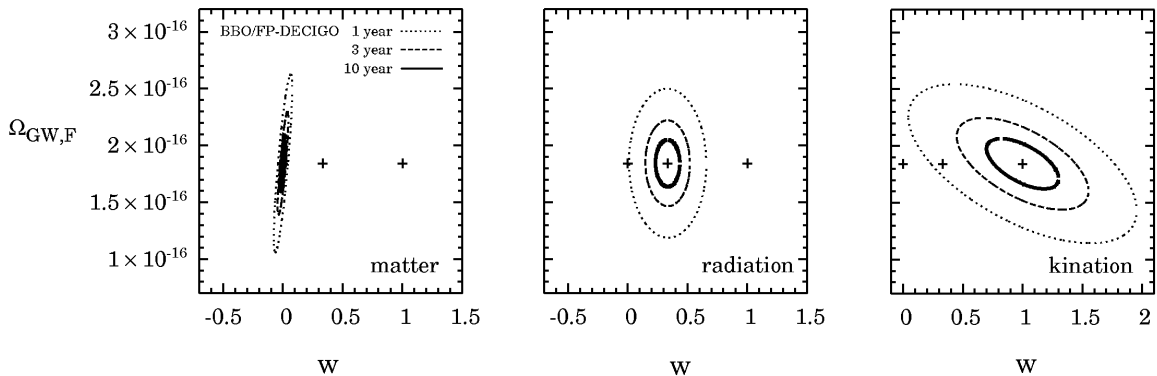}
  \caption{\label{figcontw} The $2\sigma$ confidence level contours in the
    $w-\Omega_{\rm GW,F}$ plane for 1-year (dotted), 3-year
    (dashed) and 10-year (solid) observation by BBO/FP-DECIGO.  The
    fiducial value of $\Omega_{\rm GW,F}$ is taken to be $1.84\times
    10^{-16}$ which corresponds to $r=0.1$.  The three panels
    represent a different model for the equation of state;
    $w=0$ (matter dominant, left panel), $w=1/3$ (radiation
    dominant, middle panel) and $w=1$ (kination dominant, right
    panel).  These three fiducial points are shown as cross marks in
    each panel. }
 \end{center}
\end{figure}

The fisher matrix is calculated by substituting Eq. (\ref{OmegaGW2})
into Eq. (\ref{Fisher}), taking $\Omega_{\rm GW,F}$ and $w$ as
parameters.  We investigate three fiducial models of the very early
Universe; matter-dominated ($w=0$), radiation-dominated ($w=1/3$) and
kination-dominated ($w=1$), which correspond to the frequency
dependence of $f^{-2}$, $f^0$ and $f^1$, respectively.  Here, the
normalization is taken at $F=0.203\,$Hz for BBO/FP-DECIGO and
$F=0.158\,$Hz for Ultimate-DECIGO, which is chosen as the covariance
matrix to be diagonalized for the flat spectrum \cite{Seto:2005qy} and
is almost in the middle of the sensitivity range.  Fig. \ref{figcontw}
is one example of the error contours with the amplitude chosen to be
$\Omega_{\rm GW,F}=1.84\times 10^{-16}$ corresponding to $r=0.1$.
From the figure, we see the constraint on $w$ is significantly better
in the matter-dominated case.  This is because the measurable quantity
in the experiment is not $w$ but the tilt of the spectrum,
$n_T=2(3w-1)/(3w+1)$.  In this parametrization, the value of the tilt
is more sensitive to the change of $w$ when the fiducial model is
$w\sim 0$ than when $w$ has larger value.  This makes it easier for
direct detection experiments to distinguish the model with smaller
value of $w$.

\begin{figure}
 \begin{center}
  \includegraphics[width=1.00\textwidth]{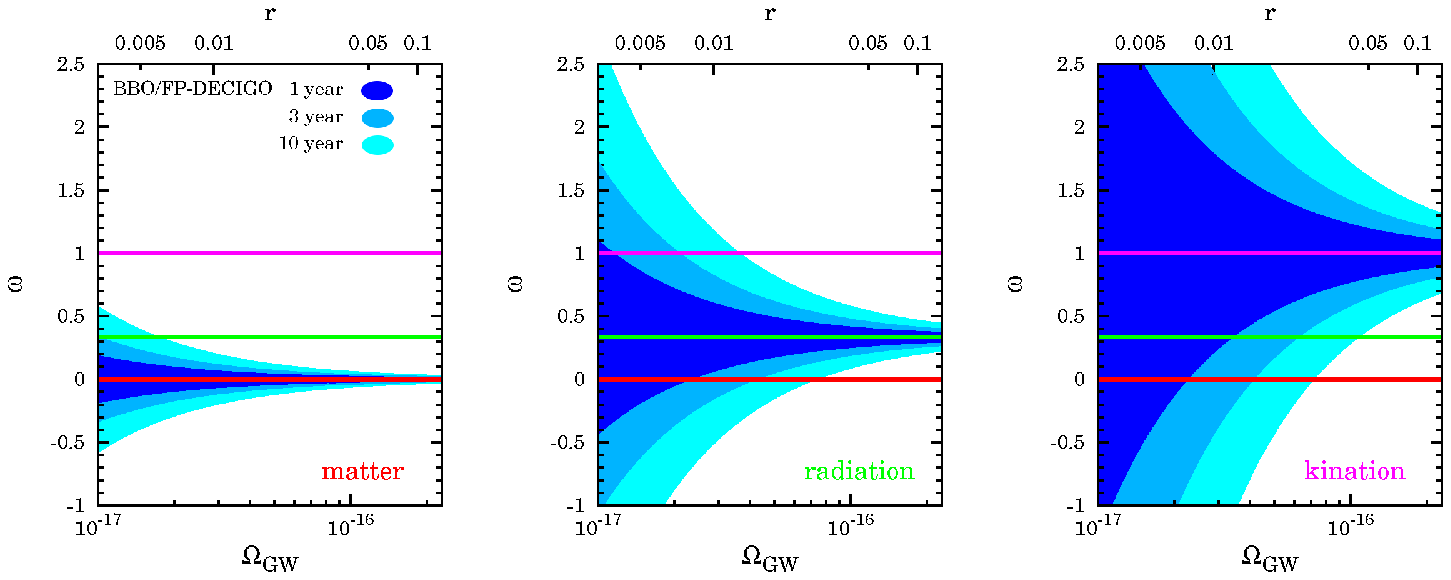}\\
  \includegraphics[width=1.00\textwidth]{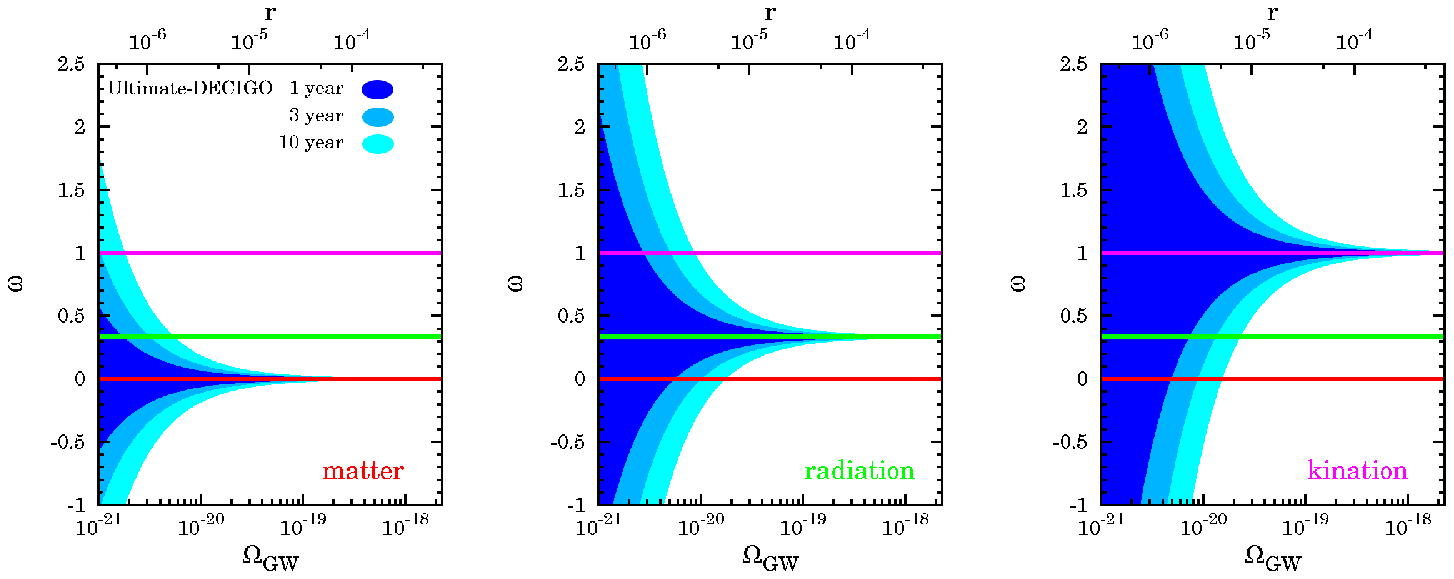}
  \caption{\label{figsigmaw} Dependence of the errors in $w$ on the
    fiducial value of the gravitational wave amplitude, $\Omega_{\rm
      GW,F}$.  The upper horizontal axis represents the values of $r$
    corresponded to $\Omega_{\rm GW,F}$ by Eq. (\ref{OmegaGW1}).  The
    solid straight lines represent values of $w$ for the matter
    ($w=0$), radiation ($w=1/3$) and kination ($w=1$)-dominated
    universe.  The blue shaded regions indicate the $1\sigma$ error
    range on each value of $w$ for 1,3 10-year observations.  Each
    three panels show the case where the fiducial model assumes the
    matter (left), radiation (middle) and kination (right)-dominated
    universe.  The upper panel is for BBO/FP-DECIGO and the lower
    panel is for Ultimate-DECIGO.}
 \end{center}
\end{figure}

Fig. \ref{figsigmaw} shows dependence on the fiducial value of
$\Omega_{\rm GW,F}$.  The marginalized error $\sigma_{w}$ is
calculated by changing $\Omega_{\rm GW,F}$ assuming the
matter-dominated, radiation-dominated and kination-dominated universe,
respectively.  Note that the vertical axis does not represent
$\sigma_{w}$, but represents $w$.  The blue shaded regions represent
the $1\sigma$ error ranges for the determination of $w=0,1/3$ and $1$,
calculated assuming 1, 3 and 10-year observation with BBO/DECIGO and
Ultimate-DECIGO.  Obviously, the errors become smaller as the
amplitude of the gravitational wave background increases, which
enables us to distinguish the value of $w$ from that of the other
models.

\section{Conclusions and discussion}

In this paper, we study prospects for direct determination of the
reheating temperature after inflation and the equation of state of the
early Universe with future gravitational wave detectors.  The
reheating temperature, $T_{\rm R}$, will be determined accurately for
$T_{\rm R} \sim 10^7$GeV by BBO/FP-DECIGO, if the tensor-to-scalar
ratio $r$ is larger than $\sim 0.05$.  Therefore, if the CMB B-mode
polarization is measured by the Planck satellite or ground-based
telescopes at the level of $r\sim 0.1$, we will have a good chance to
detect the inflationary gravitational waves at future space laser
interferometers and determine/constrain the reheating temperature.
Since $T_{\rm R}$ of $10^{6-9}$GeV is close to the upper bound from
the gravitino problem in supergravity~\cite{Kawasaki:2004yh}, the
detection of the gravitational waves will have an impact on the
supersymmetric models if the Large-Hadron Collider finds signals of
supersymmetry.  Determination of the reheating temperature is also
closely related to the origin of the ordinary matter in the present
Universe, since the amount of baryon asymmetry crucially depends on
$T_{\rm R}$ for many baryogenesis mechanisms.  For example, thermal
leptogenesis scenario needs $T_{\rm R}\gtrsim
10^9$GeV~\cite{Fukugita:1986hr}, and nonthermal leptogenesis needs
$T_{\rm R}\gtrsim 10^6$\,GeV~\cite{Asaka:1999yd,Hamaguchi:2001gw}.

Moreover, future gravitational wave detectors have a potential to
determine equation of state of the early Universe to a good accuracy.
The matter, radiation and kination dominated universe are clearly
distinguished by BBO/FP-DECIGO for $r\sim 0.1$.  It will be a
important step toward the complete understanding of the whole history
of the Universe and the nature of the inflaton itself.

As a final remark, we mention another probe of reheating.  Accurate
measurements of the CMB anisotropy are also sensitive to the reheating
temperature \cite{Martin:2010kz,Mielczarek:2010ag}.  The expansion
history of the Universe can be connected to the length of inflation
and slightly affects the values of the slow-roll parameters.  This is
reflected both in the primordial spectra of the scalar and tensor
perturbations, and is also applicable to direct detection of the
inflationary gravitational wave background.  Hence, direct detection
helps to constrain the reheating temperature even in the case where
the reheating signature is not seen in the frequency coverage of
experiments \cite{Kuroyanagi:2009br}.  Also, if there is a process
called preheating \cite{Kofman:1994rk}, gravitational waves generated
during preheating can be another probe of reheating
\cite{Khlebnikov:1997di,Easther:2006gt,GarciaBellido:2007dg}.  These
gravitational waves have a peak at the characteristic frequency,
corresponding to the comoving Hubble scale at the end of inflation.
Thus, they may provide us with complementary information on properties
of inflation and reheating.

\begin{acknowledgments}

  This work is supported in part by Grant-in-Aid for Scientific
  research from the Ministry of Education, Science, Sports, and
  Culture (MEXT), Japan, No.\ 21111006 (K.N.), No.\ 22244030 (K.N.).
  S.S. acknowledges the Japan Society for the Promotion of Science
  (JSPS) for a support through the postdoctoral fellowship for
  research abroad, No.~23-529.

\end{acknowledgments}


\end{document}